\documentclass[%
 aip,
 jmp,%
 amsmath,amssymb,
 reprint,%
]{revtex4-1}

\usepackage{graphicx}
\usepackage{dcolumn}
\usepackage{bm}

\begin{document}

\title{Folding a Small Protein Using Harmonic Linear Discriminant Analysis}

\author{Dan Mendels}
\affiliation{ Department of Chemistry and Applied Biosciences, ETH Zurich, c/o USI Campus, Via Giuseppe Buffi 13, CH-6900, Lugano, Ticino, Switzerland }
\affiliation{Facolt{\'a} di Informatica, Istituto di Scienze Computazionali, Universit{\'a} della Svizzera italiana (USI), Via Giuseppe Buffi 13, CH-6900, Lugano, Ticino, Switzerland}

\author{Giovannimaria Piccini}
\affiliation{ Department of Chemistry and Applied Biosciences, ETH Zurich, c/o USI Campus, Via Giuseppe Buffi 13, CH-6900, Lugano, Ticino, Switzerland }
\affiliation{Facolt{\'a} di Informatica, Istituto di Scienze Computazionali, Universit{\'a} della Svizzera italiana (USI), Via Giuseppe Buffi 13, CH-6900, Lugano, Ticino, Switzerland}

\author{Z. Faidon Brotzakis}
\affiliation{ Department of Chemistry and Applied Biosciences, ETH Zurich, c/o USI Campus, Via Giuseppe Buffi 13, CH-6900, Lugano, Ticino, Switzerland }
\affiliation{Facolt{\'a} di Informatica, Istituto di Scienze Computazionali, Universit{\'a} della Svizzera italiana (USI), Via Giuseppe Buffi 13, CH-6900, Lugano, Ticino, Switzerland}

\author{Yi I. Yang}
\affiliation{ Department of Chemistry and Applied Biosciences, ETH Zurich, c/o USI Campus, Via Giuseppe Buffi 13, CH-6900, Lugano, Ticino, Switzerland }
\affiliation{Facolt{\'a} di Informatica, Istituto di Scienze Computazionali, Universit{\'a} della Svizzera italiana (USI), Via Giuseppe Buffi 13, CH-6900, Lugano, Ticino, Switzerland}

\author{Michele Parrinello}
\email{parrinello@phys.chem.ethz.ch}
\affiliation{ Department of Chemistry and Applied Biosciences, ETH Zurich, c/o USI Campus, Via Giuseppe Buffi 13, CH-6900, Lugano, Ticino, Switzerland }
\affiliation{Facolt{\'a} di Informatica, Istituto di Scienze Computazionali, Universit{\'a} della Svizzera italiana (USI), Via Giuseppe Buffi 13, CH-6900, Lugano, Ticino, Switzerland}


\begin{abstract}
\noindent Many processes of scientific importance are characterized by time scales that extend far beyond the reach of standard simulation techniques. To circumvent this impediment a plethora of enhanced sampling methods has been developed. One important class of such methods relies on the application of a bias that is function of a set of collective variables specially designed for the problem under consideration. The design of good collective variables can be challenging and thereby constitutes the main bottle neck in the application of these methods. To address this problem, recently we have introduced Harmonic Linear Discriminant Analysis, a method to systematically construct collective variables. The method uses as input information on the metastable states visited during the process that is being considered, information that can be gathered in short unbiased MD simulations, to construct the collective variables as linear combinations of a set of descriptors. Here, to scale up our examination of the method's efficiency, we applied it to the folding of Chignolin in water. Interestingly, already before any biased simulations were run, the constructed one dimensional collective variable revealed much of the physics that underlies the folding process. In addition, using it in Metadynamics we were able to run simulations in which the system goes from the folded state to the unfolded one and back, where to get fully converged results we combined Metadynamics with Parallel Tempering. Finally, we examined how the collective variable performs when different sets of descriptors are used in its construction.
\end{abstract}

\maketitle

\newpage
\section*{Introduction}

Simulations of complex processes such as drug binding, protein association, protein folding, phase transitions, etc. have proven to be of great value and are a pillar of contemporary scientific investigation. However, many such processes are characterized by very long time scales which prohibit their simulation using conventional simulation techniques. Hence, to circumvent this limitation, a plethora of enhanced sampling methods has been developed over the years including replica exchange based methods such as Parallel Tempering \cite{sugita1999replica} and bias based techniques such as Umbrella Sampling \cite{Torrie1977}, Metadynamics  \cite{laio_parrinello_2002} and Variationally Enhanced Sampling \cite{valsson_parrinello_2014}. The latter category relies on the use of collective variables (CVs) which describe the most essential degrees of freedom of the processes being considered. Constructing appropriate CVs however, can be challenging and time consuming. Thus, and in light of the expected continuing increase in the complexity and size of the systems being studied, devising techniques for the systematic construction of efficient CVs is regarded as an important objective of the enhanced sampling community. Also, finding good CVs is not only a technical issue, but is a way of encoding in a compact and transparent way the essence of the process being considered.

In the effort to address this challenge, in a recent publication \cite{mendels2018collective,Piccini_Mendels2018} we have proposed a new scheme for constructing systematically viable CVs through the utilization of the supervised learning class classification paradigm, and in particular using a modification of Fisher's Linear Discriminant Analysis (LDA), termed Harmonic Linear Discriminant Analysis (HLDA) (see also ref. \cite{2018arXiv180210510S}). The LDA assumes a multivariate normal distribution of the descriptors. For this reason we also examine how deviation from multivariate normality affect the CVs efficiency. The scheme of choice requires as input only short unbiased trajectories, for each metastable state.  Using these data, HLDA can estimate the direction within an $N_d$ dimensional space of selected system descriptors upon which the projections of these sets of data are best separated. The linear combination corresponding to this direction is then utilized as the CV. 

To test its applicability, HLDA has been used in ref. \cite{mendels2018collective} in two examples taken from the realm of Materials Science and Chemistry. In both cases HLDA was found to be able to generate good CVs, leading to biased runs characterized by high frequency of transitions between the metastable states of interest and to a rapidly converging sampling. 

The application of HLDA in ref. \cite{mendels2018collective} was, however, still confined to a set of relatively simple problems. Hence, if its use (and the class classification paradigm underlying it) is to be adopted for scientific and technological problems of increasing complexity, it would need to prove effective in such systems. Here, we would like to investigate the performance of HLDA for a relatively more complex system. We consider the case of a small protein, Chignolin, for which extensive simulations on purpose build machines are available \cite{Lindorff-Larsen2011}. The procedure to determine the HLDA CVs requires the use of a convenient set of descriptors $d_{i}(R)$ that are capable of describing the initial and final states. This study will give us information on the effect of the choice of descriptors on the CVs quality. 

We show also how the HLDA CVs bring out much of the physics even before performing the simulations. Moreover, we find that employing the HLDA CVs within Metadynamics simulations enables sampling numerous folding and unfolding path-ways and that through their incorporation in Parallel Tempered Metadynamics (PTMetaD) simulations \cite{bussi2006free}, estimates for the system Free Energy Surface (FES) can be obtained.


\section*{Methods}

\subsection*{Constructing the collective variable}

To construct CVs which describe the Chignolin folding process we utilize the paradigm introduced in ref. \cite{mendels2018collective} that estimates the direction $\mathbf{W}$ in an $N_d$ dimensional descriptor space in which the projections of the unbiased distributions of the folded and unfolded states are best separated. As in ref. \cite{mendels2018collective} we do this using HLDA, a modification of Fisher's LDA. Thus, as in Fisher's LDA, the estimation of $\mathbf{W}$ is done through the maximization of the ratio between the system's so called between class $\mathbf{S}_b$ and within class $\mathbf{S}_w$ scatter matrices. Like LDA, the former is measured by the square of the distances  between the projected means, and can be written as

\begin{equation}
\label{mean_transf_mat}
\mathbf{W}^T \mathbf{S}_b \mathbf{W}
\end{equation}

\noindent
with

\begin{equation}
\label{between_class}
\mathbf{S}_b = \left( \boldsymbol{\mu}_A - \boldsymbol{\mu}_B \right)\left( \boldsymbol{\mu}_A - \boldsymbol{\mu}_B \right)^T
\end{equation}

\noindent
where $\boldsymbol{\mu}_{A,B}$ are the expectation values of the two metastable states. In contrast to LDA, in which the within class matrix is estimated using the average of the two metastable states multivariate variances $\boldsymbol{\Sigma}_{A,B}$, here it is estimated from the spreads harmonic average  

\begin{equation}
\label{cov_transf_mat}
\mathbf{W}^T \mathbf{S}_w \mathbf{W}
\end{equation}

\noindent
with 

\begin{equation}
\label{harmonic_mean}
\mathbf{S}_w = \frac{1}{\frac{1}{\boldsymbol{\Sigma}_A} + \frac{1}{\boldsymbol{\Sigma}_B}}.
\end{equation}

\noindent
The HLDA objective function which has the form of a Rayleigh ratio
 
\begin{equation}
\label{fisher_ration}
\mathcal{J(\mathbf{W})} = \frac{\mathbf{W}^T \mathbf{S}_b \mathbf{W}}{\mathbf{W}^T \mathbf{S}_w \mathbf{W}}
\end{equation}

\noindent
is then maximized by

\begin{equation}
\label{maximizer}
\mathbf{W}^* = \mathbf{S}_w^{-1} \left( \boldsymbol{\mu}_A - \boldsymbol{\mu}_B \right).
\end{equation}

\noindent
which in turn yields the HLDA CV

\begin{equation}
\label{maximizer_harm}
s_{HLDA}(\mathbf{R}) = \left( \boldsymbol{\mu}_A - \boldsymbol{\mu}_B \right)^T \left( \frac{1}{\boldsymbol{\Sigma}_A} + \frac{1}{\boldsymbol{\Sigma}_B} \right)   \mathbf{d}(\mathbf{R}).
\end{equation}

\subsection*{Computational Details}

Simulations of Chignolin (sequence TYR-TYR-ASP-PRO-GLU-THR-GLY-THR-TRP-TYR) were conducted using GROMACS 5.1.2 \cite{berendsen1995gromacs,abraham2015gromacs} and the PLUMED 2.4 plugin \cite{tribello_2014}. The CHARMM22* force field \cite{piana2011robust} and the three-site transferable inter-molecular potential (TIP3P) water model \cite{jorgensen1983comparison} were used to make direct comparisons with ref. \cite{lindorff2011fast}. ASP and GLU residues were simulated in their charged states, as were the N- and C-terminal amino acids. A time step of 2 fs was used for all systems and a constant temperature of 340 K (in agreement with ref. \cite{lindorff2011fast}) was maintained by the velocity rescaling thermostat of Bussi et al. \cite{bussi2007canonical}. All bonds involving H atoms were constrained with the linear constraint solver (LINCS) algorithm \cite{hess1997lincs}.  Electrostatic interactions were calculated with the particle mesh Ewald scheme \cite{essmann1995smooth} and 1 nm cutoff was applied to all non-bonded interactions. Runs were all conducted with a box containing 1649 water molecules, along with 2 sodium ions to neutralize the system. Parallel Tempering simulations were run with 40 replicas, each at a different temperature. The replica with the lowest temperature was set to be at T=340 K while the rest of the temperatures were arranged in a geometrical series with a factor $a=1.011$. (Details regarding the utilized descriptor sets can be found in the Supplementary Information).

\section*{Results}

In accordance with the basic requirements of the HLDA approach, we began the study by acquiring two unbiased trajectories spanning roughly $2 \mu s$ each in the system's folded and unfolded states. The first step in applying HLDA requires the selection of a set of system descriptors that can be instrumental in describing the folding and unfolding of the mini-protein. Ideally, since this step should not require an expert's understanding of the system we proceeded by selecting fairly naively three different sets of descriptors to observe how in the present context this selection can influence the outcome of the method implementation. Here, we also chose to asses the HLDA ability to perform beyond its strict theoretical limitations, namely that the descriptor unbiased fluctuations are normal in form.

Thus, the first set $D_1$ consisted of 12 distances between different atomic sites on the protein. Six of these distances were selected between atoms situated on the backbone, while six more where taken between atoms situated on the protein's side chains. The second set of descriptors $D_2$ consisted of 6 contacts placed on the protein's backbone while the third $D_3$ consisted of $\alpha \beta$ functions corresponding to the protein's backbone dihedral angles, i.e. $\alpha \beta=\frac{1}{2}(1+cos(\phi_i))$ with $i=1..18$. (For a detailed list of the atom pairs used for the construction of  $D_1$ and $D_2$ and the contacts parameters see the SI). The two covariance matrices $\boldsymbol{\Sigma_{f}}$, $\boldsymbol{\Sigma_{u}}$ and two mean vectors $\boldsymbol{\mu_{f}}$, $\boldsymbol{\mu_{u}}$ corresponding to the folded and unfolded states respectively, were thus constructed for each of the descriptor sets.

\begin{figure}
	\centering
    \includegraphics[width=1\columnwidth]{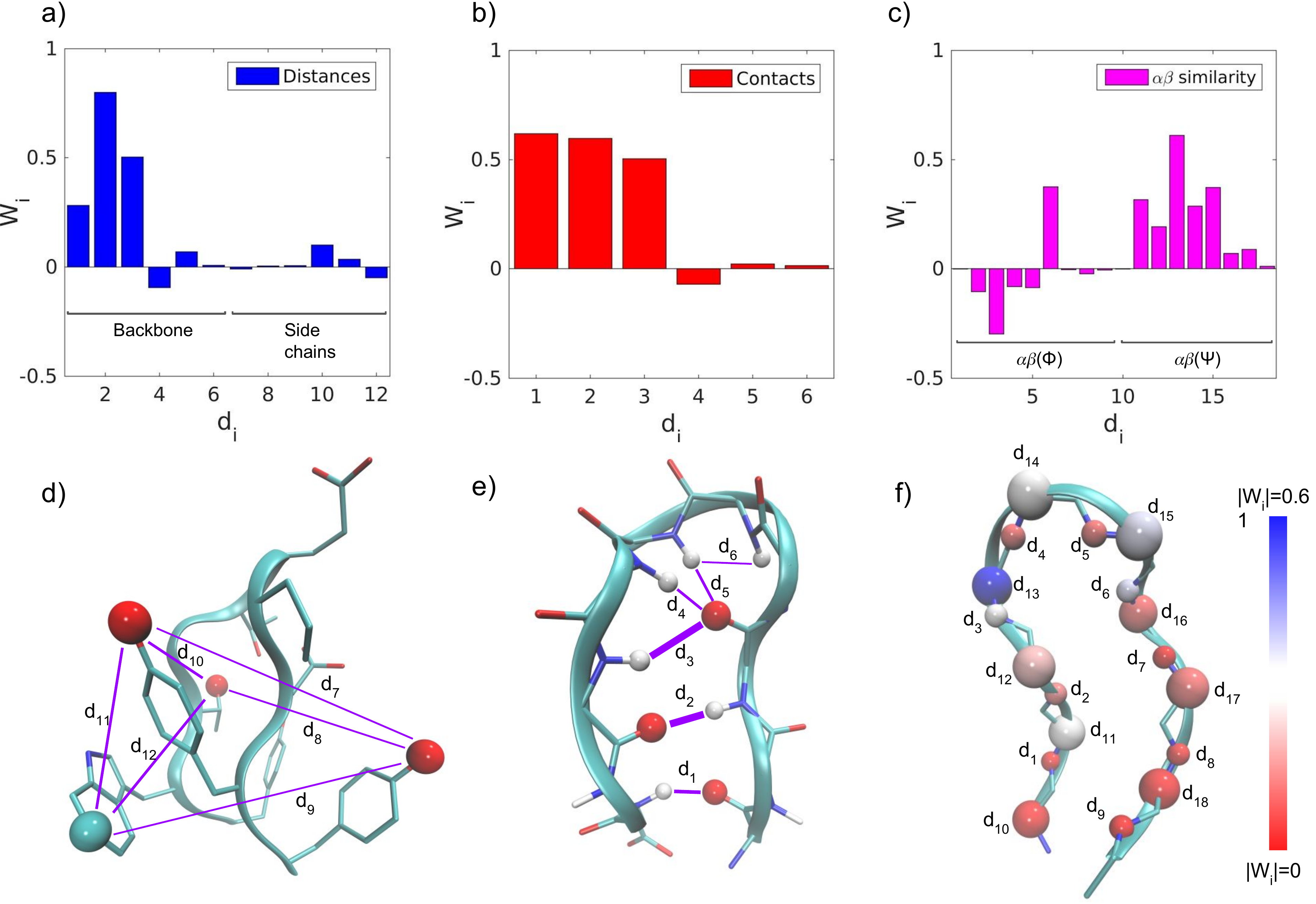}
    \caption{The weights assigned by HLDA to each of the descriptors for a) $D_1$ b) $D_2$ and c) $D_3$. d) Illustration of the distances between side-chain sites utilized in the set $D_1$. e) Illustration of the distances and contacts used between backbone sites in the sets $D_1$ and $D_2$ (the line thicknesses are set to indicate the descriptors' importance in both sets). f) Illustration of the descriptor set $D_3$. Sphere colors correspond to the absolute values of the weights assigned to each of the utilized backbone dihedral angles. Small spheres represent the $\Phi$ angles whereas large ones the $\Psi$ angles.  
    }
    \label{fig:Illustration_descriptors}
\end{figure}

Using this information and applying HLDA we could next obtain an estimation of the hyper planes that best separated the unbiased distributions corresponding to the folded and unfolded states within the space spanned by each set of descriptors. Concomitantly, the sought after CVs were obtained using Eq. \ref{maximizer_harm}. The weights of the HLDA CVs attained for each of the descriptor sets are plotted in Fig. \ref{fig:Illustration_descriptors} along with an illustrations of the utilized descriptors. 

Interestingly, analysis of the weight distributions reveals much valuable information about the system showing that the main features of the folding process are encoded in the CVs themselves. Thus, we find that for both the sets $D_1$ and $D_2$ most of the weight is assigned to the descriptors $d_1$, $d_2$ and $d_3$ which correspond to the distances and contacts between facing amino acids located away from the backbone beta-turn. Similarly, in both $D_1$ and $D_2$ the distances/contacts located in the beta-turn are found to be comparably less important, alluding to the fact that in the unfolded state the beta-turn is intermittently formed. In the case of $D_1$ the distances between the side chains are assigned lower weights as well. Nevertheless, for the distances $d_{10}$, $d_{11}$ and $d_{12}$ non negligible weight is assigned, reflecting the associated side-chain's contact formation in the folded state, due to their hydrophobic nature.

Inspecting the weight distribution obtained for $D_3$ reveals interesting trends as well. Thus, one can observe that by and large the higher weights of the CV are assigned to $\alpha\beta$s of the backbone dihedral angles situated in and near the backbone beta-turn, reflecting these angles' importance in the folding process. In addition, we find that in comparison it is the $\alpha\beta(\Psi)$ that attain higher weights. Inspection here shows that while the $\alpha\beta(\Phi)$ fluctuations do not change much between the folded and unfolded states, clearly configurational changes of $\alpha\beta(\Psi)$ between the folded and unfolded states are present. Another interesting feature of the CV weight distribution is that with the exception of $d_6$, the weights of $\alpha\beta(\Psi)$ and $\alpha\beta(\Phi)$ tend to be in anti-phase. Here, inspection shows that this results from the correlation between the  $\alpha\beta(\Psi)$ and $\alpha\beta(\Phi)$ fluctuations in the $\alpha$-helical basin that is visited in the unfolded state. Moreover, we attribute the single positive value of $d_6$ (which corresponds to the Glycine $\Phi$) to the fact that unlike the other $\Phi$ backbone dihedral angles it can also assume a left helix conformation \cite{Lovell2003}. Finally, examination of $d_{13}$, the descriptor assigned with the largest weight, shows that it corresponds to the $\alpha\beta(\Psi)$ of the Proline amino acid, coinciding with the fact that such angles are associated with a relatively high energetic rotational barrier \cite{KANG2004135}.

\maketitle

\subsection*{Biased Simulations}

\begin{figure}
	\centering
    \includegraphics[width=0.47 \columnwidth]{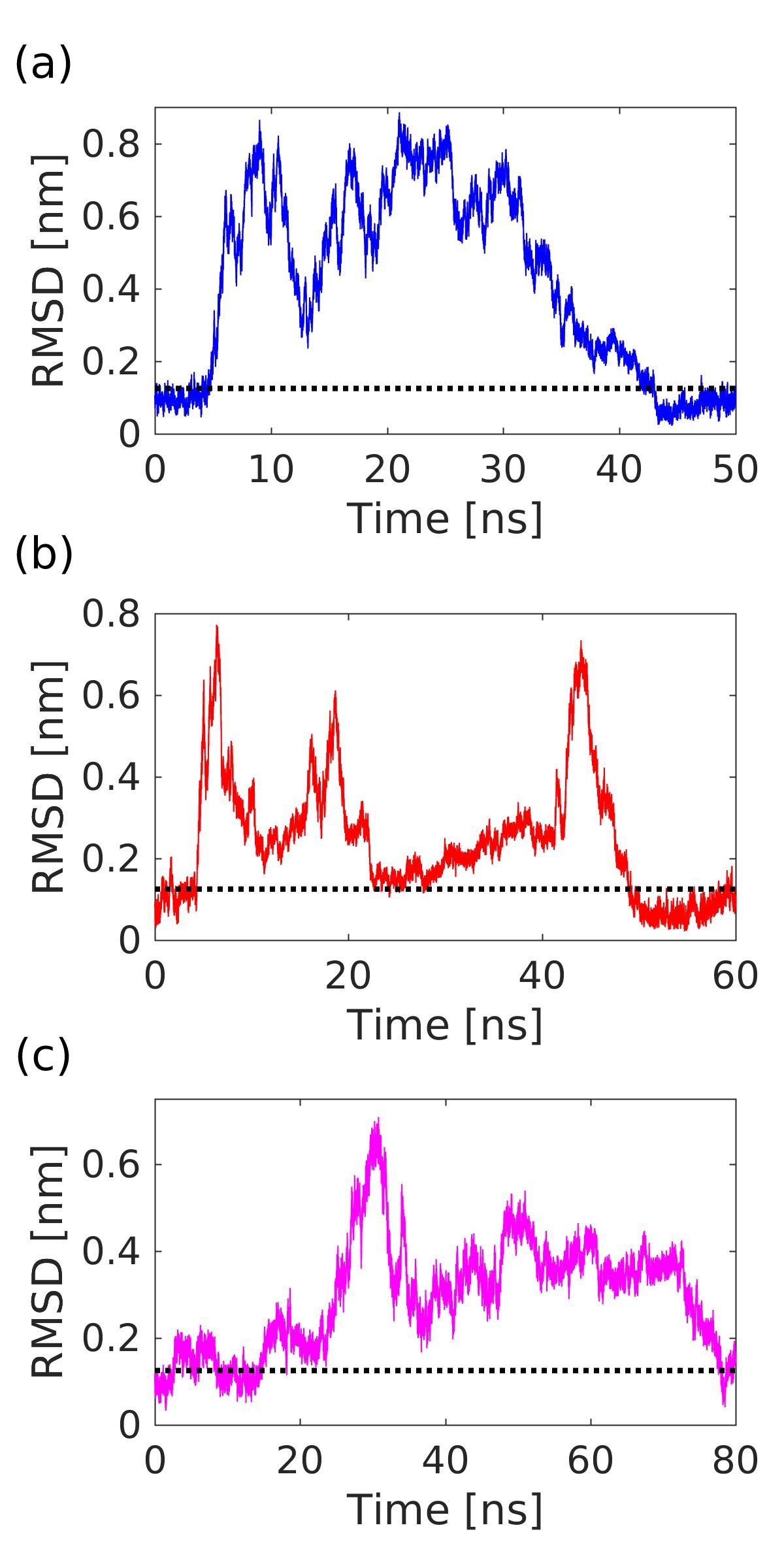}   
    \caption{Excerpts from Metadynamics simulations utilizing the CVs generated with HLDA, applied on the descriptor sets (Top) $D_1$, (Middle) $D_2$ and (Bottom) $D_3$.
    }
    \label{fig:Paths_discovery}
\end{figure}

With the CVs at hand we could next launch Metadynamics simulations with the objective of sampling the system phase space. Monitoring these simulations, we found that several folding events could be observed with $D_1$ and $D_2$ taking the lead in the transition frequencies. Thus, with the little initial information with which we commenced, an assortment of folding and unfolding pathways could be harvested, thereby shedding light on the mechanisms underlying these events. Fig. \ref{fig:Paths_discovery} presents segments of three simulations, each run with a CV generated by a different descriptor set, showing the $C\alpha$ RMSD of the protein with respect to it's folded crystal structure as function of Metadynamics simulations time. All three segments exhibit both folding and unfolding events. 

\begin{figure}
	\centering
    \includegraphics[width=0.52\columnwidth]{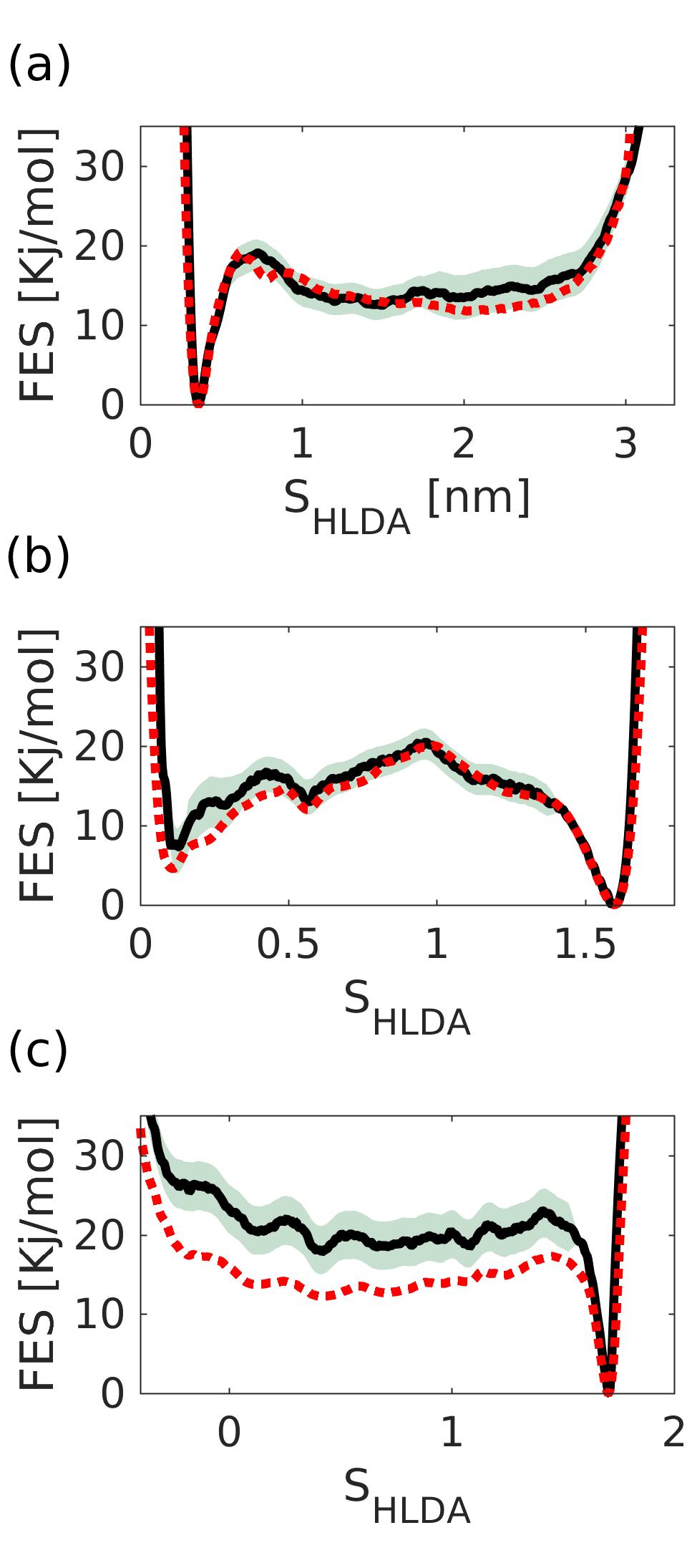}   
    \caption{Free-energy profiles at $T=340 K$ obtained from PTMetaD simulations along the $S_{HLDA}$ CV constructed using the descriptor set (a) $D_1$, (b) $D_2$, (c) $D_3$. Shaded areas indicate the fluctuations in time of the FES curves during convergence. Dotted lines, represent the FES profiles obtained via a histogram analysis on the $100 \mu s$ unbiased trajectories taken from the D.E. Shaw database.
    }
    \label{fig:FES_HLDA_PT_compare}
\end{figure}

Despite observing even multiple transitions between the folded and unfolded states in some of the simulations, attaining estimates of converged FES using Metadynamics alone was not possible. Observing the simulations dynamics we found this to be caused by the system's phase space intricate multidimensional nature with a profusion of kinetic bottlenecks and free energy barriers. Thus, to circumvent this impediment we resorted to the utilization of Parallel Tempering Metadynamics which has previously been shown to be very effective for such problems. By running such simulations, now with the HLDA generated CVs, we could observe that for all three sets of descriptors effective sampling of the relevant system phase space was achieved. Moreover, estimates of the system FES could be obtained using the Well tempered Metadynamics (WTMD) \cite{Barducci2008} relation Eq. \ref{Eq:V_to_FES}

\begin{equation} \label{Eq:V_to_FES}
    F(\boldsymbol{s})=-\Big( \frac{\gamma}{\gamma-1} \Big) V(\boldsymbol{s}).
\end{equation}

\noindent where $\gamma$ is the WTMD bias factor and $V(s)$ is the simulation bias. Fig. \ref{fig:FES_HLDA_PT_compare} presents the FES obtained from three different simulations using the CVs generated by the three different sets of descriptors. For the sake of comparison, the corresponding FES obtained using a histogram analysis on a $100 \mu s$ unbiased simulations taken from the D.E. Shaw data bank \cite{lindorff2011fast} is presented as well. As can be seen in all three cases reasonable estimates of the FES could be obtained. However, the results obtained using the sets $D_1$ and $D_2$ are markedly more accurate. Additionally, differences in the convergence times between the different simulations were observed, namely $t_{conv}(D_1)\approx25 ns$, $t_{conv}(D_2)\approx35 ns$ and $t_{conv}(D_3)\approx45 ns$, where we defined convergence when the calculated FES fluctuations as function of simulation time reached their minimal amplitudes around their final average result.

\begin{figure}
	\centering
    \includegraphics[width=1.0\columnwidth]{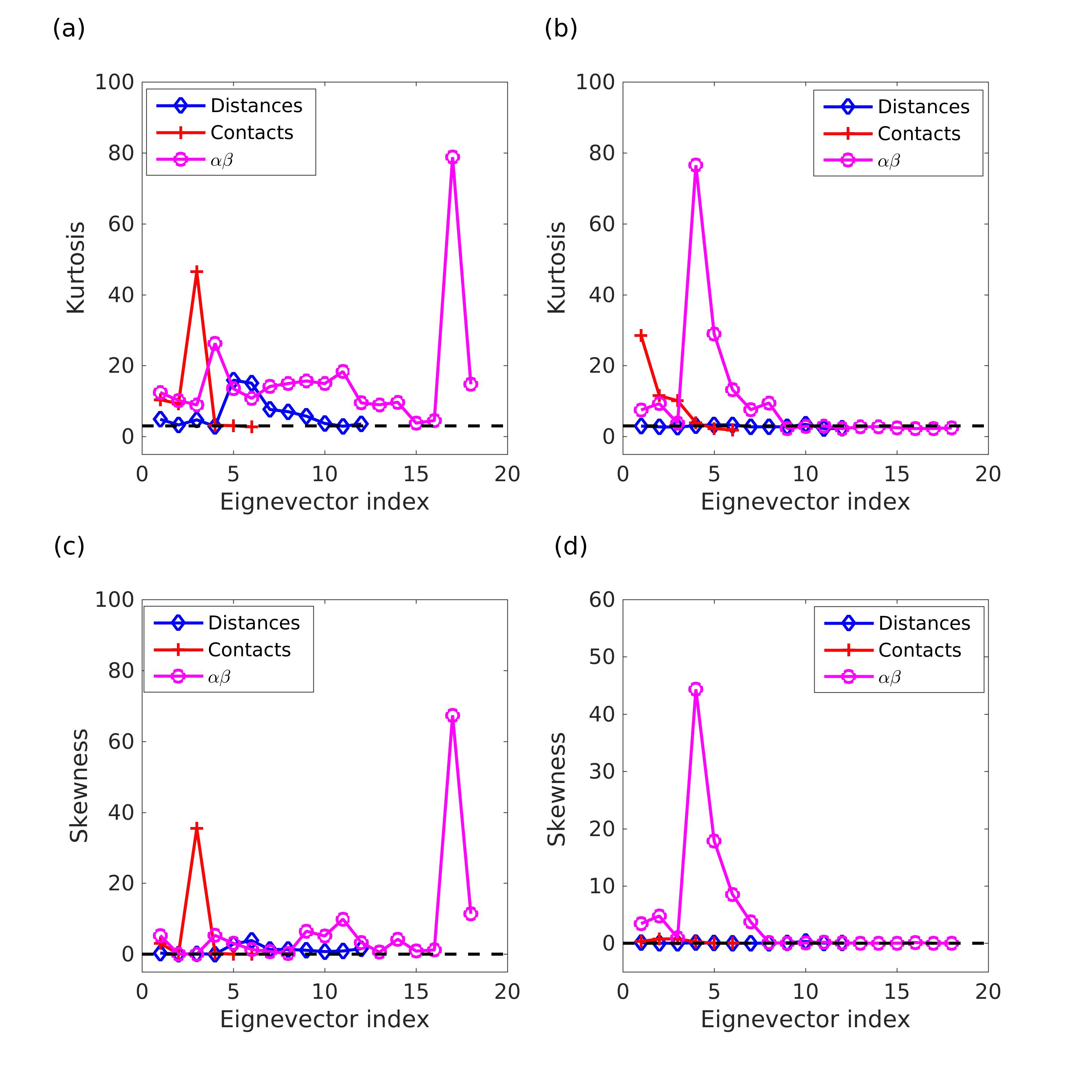}   
    \caption{Kurtosis and skewness of the unbiased distributions in the folded (left) and unfolded (right) states of the covariance matrices eigenvectors corresponding to the three utilized sets of descriptors. The expected values for perfect multinormal distributions are indicated by the dashed lines.  
    }
    \label{fig:Gaussinaity}
\end{figure}

 One probable reason for the differences in performance between the different CVs, is the extent to which their underlying descriptors unbiased fluctuations deviate from multivariate normality. To asses the possible influence of such deviations on the CVs quality we thus computed the Kurtosis and Skewness \cite{mardia1975assessment} of the covariance matrices, $\boldsymbol{\Sigma}_{f,u}$, eigenvectors. Fig. \ref{fig:Gaussinaity} presents these data for each of the descriptor sets in both the folded and unfolded states. As can be seen the values obtained for $D_1$ largely match those that correspond to perfect normal distributions (indicated by dashed lines). In slight contrast, the set $D_2$ can be seen to exhibit larger deviations from normality, while the set $D_3$ seems to exhibit the most and largest instances of such deviations. Observation of these data thus indicates that while HLDA seems to be forgiving when applied to data which is not strictly multinormal, it is likely that a price is to be paid in their quality as deviations increase.


\section*{Conclusions}

As the size and complexity of systems simulated using molecular dynamics increases, the need for systematic ways of constructing viable CVs for these systems which do not require an expert's knowledge is becoming more evident. In the present study we have applied HLDA, a recently developed modification of LDA to develop one dimensional CVs for the folding problem of Chignolin. In doing so we have found that given a naive selection of descriptors the method is able to generate CVs that, when biased, are able to drive the system back and forth between the system's folded and unfolded states. In addition, we have found that incorporating these CVs in PTMetaD can enable obtaining of good estimates of the systems FES. In both cases we found that some deviation from multivariate normality is tolerated by HLDA, yet increasing the amount of deviation may lead to a reduction in the constructed CVs' quality. Finally, we found that within the weight distributions of the calculated HLDA CVs themselves reside abundant useful information and physical insight about the process being studied. We thus conclude that the present study suggests that HLDA can be applied to increasingly more complex systems for the systematic construction of CVs, a path which we wish to continue and explore in the near future.

\begin{acknowledgments}
 We acknowledge D. E. Shaw Research for sharing data from the simulations of chignolin. This research was supported by the European Union Grant No. ERC-2014-AdG-670227/VARMET. Calculations were carried out on the M\"{o}nch cluster at the Swiss National Supercomputing Center (CSCS).
\end{acknowledgments}

\newpage
\bibliographystyle{ieeetr}
\bibliography{library}

\end{document}